\documentclass[aps,pre,twocolumn,epsf,graphicx]{revtex4}
\usepackage{amsfonts}
\usepackage{amsmath}
\usepackage{amssymb}

\def\pd2x{{\partial^2 \over \partial x^2}}

\usepackage{epsfig,latexsym}

\newcommand \bew {\begin{widetext}}
\newcommand \enw {\end{widetext}}

\begin{document}

\title{\bf\noindent The  statistical mechanics of traveling salesman 
type problems} 

\author{David S. Dean$^{(1,2)}$, David Lancaster$^{(3)}$ 
and Satya N. Majumdar$^{(2,4)}$}

\affiliation{
(1) DAMTP, CMS, University of Cambridge, Cambridge, CB3 0WA, UK \\
(2) Laboratoire de Physique Th\'eorique,  UMR CNRS 5152, IRSAMC, Universit\'e 
Paul Sabatier, 118 route de Narbonne, 31062 Toulouse Cedex 04, France\\
(3) Harrow School of Computer Science, University of Westminster, 
Harrow, HA1 3TP, UK \\
 (4) Laboratoire de Physique Th\'eorique et Mod\`eles Statistiques, UMR 8626,
Universit\'e Paris Sud, B\^at 100, 91045 Orsay Cedex, France
}
\date{3 November 2004}
\begin{abstract}
We study the finite temperature statistical mechanics of Hamiltonian 
paths between a set of $N$ quenched randomly distributed points in 
a finite domain $\cal D$. The energy of the path is a function of the 
distance between  neighboring points on the path, an example is the 
traveling salesman problem where the energy is the total distance 
between neighboring points on the path. We show how the system can be 
analyzed in the limit of large $N$ without using the replica method. 
\end{abstract}

\maketitle
\vspace{.2cm} \pagenumbering{arabic}

%%%%%%%%%%%%%%%%%%%%%%%%%%%%%%%%%%%%%%%%%%%%%%%%
In this Letter we consider a system where $N$ points with quenched
positions $x_1^{(q)},\ x_2^{(q)}\cdots x_N^{(q)}$ are independently 
distributed on a finite domain $\cal D$ with a  probability density function
$p_q(x)$. In general, the domain $\cal D$ is multidimensional
and the points $x_i^{(q)}$ are vectors in the corresponding Euclidean space.
Inside the domain $\cal D$ we consider a polymer chain composed of
$N$ monomers whose positions are denoted by $x_1$, $x_2$, $\dots$, $x_N$.
Each monomer $x_i$ is attached to one of the quenched sites $x_i^{(q)}$
and only one monomer can be attached to each site.The state of the 
polymer is described by a permutation $\sigma \in \Sigma_N$ 
where $\Sigma_N$ is the group of permutations of $N$ objects. The 
position of the $i^{th}$ monomer may thus be written as 
$x_i = x^{(q)}_{\sigma(i)}$. The Hamiltonian for the system is given by
\begin{equation}
H(\sigma) = \sum_{i=1}^N V\left(x^{(q)}_{\sigma(i)} -x^{(q)}_{\sigma(i-1)}
\right).
\end{equation}
Here $V$ is the interaction between neighboring monomers on the polymer chain.
For convenience the chain is taken to be closed, 
thus we take the periodic boundary condition
$x_{0} = x_N$. 

A physical realization of this system is
one where the $x^{(q)}_{i}$ are impurities where the monomers of a 
polymer loop are pinned. The potential  $V$ represents effective interaction
between neighboring monomers on the chain. 
For instance $V(x) = \lambda x^2/2$ corresponds to the Rouse model
of a polymer chain\cite{Rouse}. Another application of this model 
occurs in combinatorial optimization: 
if one takes $V(x)$ to be the norm, or distance, of the vector $x$ 
then $H(\sigma)$ is the total distance covered by a path which visits each 
site $ x^{(q)}_i$ exactly once. The  problem of finding $\sigma^*$ which 
minimizes $H(\sigma)$ is known as the traveling salesman problem (TSP)
\cite{mtsp}. If one takes
$V(x)= |x|$ (the Euclidean norm) then the combinatorial optimization 
problem is called the Euclidean TSP \cite{tspe}. The choice 
$V(x)= -|x|$ means that one is looking for the longest path, this is the
so called Maximal TSP or taxicab rip-off \cite{mtsp}. 
The analogy with the physical
polymer system is particularly useful. First one may use the techniques 
of simulated annealing \cite{siman} to search for the optimal path 
$\sigma^*$ by introducing a temperature in the system and then incorporating 
a dynamics respecting  detailed balance. The optimal path corresponds 
to the ground state or zero-temperature energy $E_{GS}$ of $H$. By 
slowly cooling the system one may  find this state, though in the presence of 
many local minima or metastable states this procedure may not be 
efficient. Secondly, and this is the approach taken here,
one may try to determine the full temperature dependence of the
free energy averaged over the disorder ensemble and then take the
zero temperature limit to determine the average value $E_{GS}$ 
\cite{mepavi,fuan,mamoze}. Depending on the form of $V$, $E_{GS}$ 
may not be extensive in $N$. In such cases, for example for  
the Euclidean TSP, the temperature needs to be rescaled with $N$ in order
to have an extensive ground state energy \cite{mepavi,tspps1}.
In this letter we study the problem with a fixed domain size and 
no rescaling of the temperature with $N$. The TSP is often studied in
this form and in the thermodynamic limit the ground state energy 
per site is strictly zero. 

The main technical problem associated with the second approach above 
is to perform the average over the quenched disorder.The replica 
method and cavity methods have been used previously to perform the quenched
average for the TSP \cite{tspps1,tspps2,tspcav,pema1,pema2}, 
notably in the random link version of the problem where 
the distances between sites  $i$ and $j$ are assumed to be independent.
This `random link' assumption considerably facilitates carrying out the 
disorder average
within the replica/cavity formalism. However, in the Euclidean
TSP problem the distances between the sites are evidently correlated
(for instance the triangle inequality must be respected). 
In this Letter we present an exact approach that (i) does not 
require the use of the replica method,  (ii) fully takes 
into account the correlations between the distances and
(iii) moreover provides us with exact asymptotic results at all temperatures. 
 
The canonical partition function
is given by the sum over all permutations 
\begin{equation}
Z_N = {1\over N!}\sum_{\sigma\in {\Sigma_N}} \exp\left(-\beta H(\sigma)\right).
\label{eqpart}
\end{equation}
Since the number of permutations grows as $N!$, the  entropy is of order
$N \ln N$ and we  insert a factor of $1/N!$ to absorb it.
To simplify our formulas we take unit domain size.

We define the density of quenched sites on ${\cal D}$ as
\begin{equation}
\rho_q(x) = {1\over N}\sum_{i=1}^N \delta(x-x_i^{(q)}).
\end{equation}
The partition function $Z_N$ 
only depends on $\rho_q(x)$. If $x_i$ is the site visited by the monomer 
$i$ then the partition function of the permutation problem can be written
up to constant factors as
\begin{eqnarray}
Z_N = {1\over N!}\int \prod_{i=1}^{N} 
dx_i \left[ \prod_x 
\delta\left(N \rho_q(x) - \sum_i \delta(x-x_i)\right)\right]\nonumber \\
 \exp\left(-\beta\sum_i V(x_{i}-x_{i-1})   \right).
\end{eqnarray}
The delta function constraint above ensures that the monomers $x_i$
have the same density as the quenched points and thus visit only
the quenched points and with the correct degeneracy. 
Using a Fourier representation of the functional
constraint gives, up to temperature independent constants,
\begin{equation}
Z_N = \int d[\mu] \exp\left(N\int dx  \mu(x)\rho_q(x)\right)
{\cal Z}_N.
\label{eqzq}
\end{equation}
The object ${\cal Z}_N$ is  the annealed
partition function for a {\em free} ring polymer whose monomers can 
attain any point in ${\cal D}$ but with an $x$ dependent 
chemical potential. 
It is defined as:
\begin{equation}
{\cal Z}_N = \int \prod^N_{i=1}dx_i\ 
\exp\left(-\beta \sum_{i=1}^{N} V(x_i-x_{i-1})  
-\sum_{i=0}^N \mu(x_i)\right).
\label{eqcalz}
\end{equation}
The partition function  ${\cal Z}_N$ may be evaluated by operator techniques:
${\cal Z}_N = {\rm Tr} \ T^N$ where $T$ is the symmetric operator
\begin{equation}
T(x,y) = \exp\left(-{\mu(x)\over 2}  -{\mu(y)\over 2} -\beta V(x-y)
\right).
\end{equation} 
The full partition function Eq. (\ref{eqzq}) can be evaluated by the saddle
point method in the limit where $N\to \infty$ keeping the size 
of $\cal D$ fixed. 
The saddle point equation is
\begin{equation}
\rho_q(x) = -{1\over N}{\delta \ln {\cal Z}_N \over \delta \mu(x)}
= {1\over N}\langle\sum_{i=1}^N
\delta(x-x_i)\rangle
 = p_a(x) .
\end{equation}
where the above expectation, is for the 
system with partition function ${\cal Z}_N$ defined in Eq. (\ref{eqcalz}),
so $p_a(x)$ is the, annealed, density of points (monomers) for the 
free ring polymer.
Physically this approach can be thought of as choosing a site dependent 
chemical potential $\mu$ which fixes the density of the
annealed calculation to be the same as that of the quenched one,
{\em i.e.} so that $\rho_q(x) = p_a(x)$.
The approach has some similarity with the constrained
annealing approximation \cite{morita,kuhn}, 
but fixes the whole distribution rather than individual moments.

The  ground state eigenfunction, corresponding to the 
maximal eigenvalue $\lambda_0$ of $T$ obeys 
\begin{equation}
f^{(q)}_0(x) = \lambda_0^{-1} \int dy\ T(x,y)f^{(q)}_0(y),
\label{eqfq}
\end{equation}
and for  large $N$ the annealed density of points in 
$\cal D$ is given by
\begin{equation}
p_a(x) 
= -{\delta \ln(\lambda_0) \over \delta \mu(x)}
= \left[f^{(q)}_0(x)\right]^2.
\end{equation}
Substituting this into the saddle point equation
and writing $\exp(-{\mu(x)\over 2}) = 
\sqrt{\rho_q(x)}/s_{\lambda_0}(x)$ we find that
$s_{\lambda_0}(x)$ obeys
\begin{equation}
s_{\lambda_0}(x) = \lambda_0^{-1} \int dy\ \exp\left(-\beta V(x-y)\right)
{\rho_q(y) \over s_{\lambda_0}(y)}. \label{eqsx}
\end{equation}
Considering the case of  a  uniform distribution of the points, 
$x_i^{(q)}$, and substituting back into the action we obtain
\begin{equation}
-{\beta F_N\over N} = 2\int dx\ \ln\left(s_{\lambda_0}(x)\right) \   
+ \ln(\lambda_0)
+{\rm terms\ indep.\ of\ }\beta . 
\end{equation}
From  Eq. (\ref{eqsx}) we see that there is a family of 
solutions $\lbrace s_{\lambda_0}(x), \lambda_0\rbrace$ which are related by 
$s_{\lambda_0} = a^{1/2}s_{a\lambda_0}$, for $a>0$ and in addition these 
solutions all have the same action. This apparent zero mode is an artifact
introduced by the fact that the constraint $N = N\int dx\  \rho_q = \int dx\  
\sum_i\delta(x-x_i)$ is automatically satisfied. 
Thus we may chose $\lambda_0 = 1$, leading
to our final  result for the average energy per site
\begin{eqnarray}
\epsilon &=& -2{\partial \over \partial \beta}\left[ \int dx\ 
\ln\left(s(x)\right)\right] \nonumber \\
         &=& \int dx\ dy\  {V(x-y) \exp\left(-\beta V(x-y)\right)\over
s(x) s(y)},\label{eqsf}
\end{eqnarray}
where $s$ obeys
\begin{equation}
s(x) =\int dy\ {\exp\left(-\beta V(x-y)\right)\over 
s(y)}. \label{eqsff}
\end{equation}  

In general Eq. (\ref{eqsff}) can be solved by an iterative 
numerical procedure.
The average energies obtained in this way for the 
one dimensional TSP on the unit interval $[0,1]$ and the conventional 
two dimensional TSP on the square domain $[0,1]^2$ are shown 
as  continuous lines in Fig. (1). 
To test these predictions
we have carried out Monte Carlo simulations of this TSP for system 
sizes of $N=2000$ and compared the average energy measured after equilibrating 
the system over $10^6$ Monte Carlo steps and measuring the average energy 
over a subsequent $10^6$ Monte Carlo steps. 
The basic move in the dynamics was a transposition of a pair of 
points in the permutation, and the acceptance criterion was the 
Metropolis rule.  The simulation points are also shown as solid 
circles on Fig. (1) and  we see that for all temperatures the 
agreement with the theoretical prediction is excellent. 

\begin{figure}
\epsfxsize=0.8\hsize \epsfbox{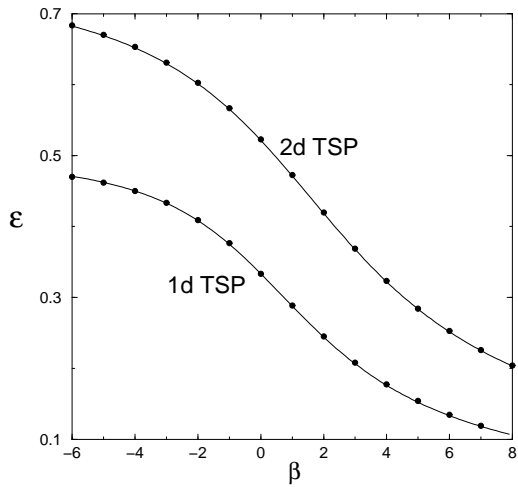}
\caption{Theoretical prediction for the average energy $\epsilon$ for 
one and two dimensional TSP as a function 
of $\beta$ (solid line) compared with the  
Monte Carlo simulations (solid circles). Negative $\beta$
corresponds to the maximal TSP problem.
Error bars based on 20 realizations of the quenched points
are smaller than the symbol sizes.}
\label{fig:2DTSP}
\end{figure}   

For the one dimensional ring and two dimensional torus or sphere, 
where the Euclidean distance is taken to be 
the shortest way round the domain, Eq. (\ref{eqsff}) admits a 
constant solution. This indicates
that in these cases, the annealed approximation is exact without
the need for any chemical potential to force the density
of the annealed points to be uniform. In the case of the
ring we find
\begin{equation}
\epsilon = {1\over \beta} - {1\over 2(e^{\beta/2} - 1)},
\label{eqring}
\end{equation}
for both positive and negative $\beta$. This result, and
the corresponding one for the torus and sphere are confirmed by Monte
Carlo simulations, and moreover support the correctness
of the iterative procedure used to numerically solve
Eq. (\ref{eqsff}) \cite{inprep}.

In one dimension the Euclidean TSP amounts to a sorting problem, 
but even when the quenched points are regularly spaced, 
no simple expression for the partition function is known \cite{djl}. 
For the unit interval ${\cal D}=[0,1]$ 
we have obtained the low temperature expansion
\begin{eqnarray}
\epsilon &=& {1\over \beta} - {\pi^2 -6\ln^2(2)\over 6\beta^2} \ \; \beta \to \infty, \nonumber\\
\epsilon &=& {1\over 2} - {\pi^2\over {6\beta^2}} \ \; \beta \to -\infty,
\end{eqnarray}
which agrees with our numerical results.
We have also considered the one-dimensional potentials
$V(x) = x^2$ and $V(x) = -\ln(|x|)$, the latter was only considered at 
positive temperature as it is ill defined at negative temperature. 
In all these cases, the comparison with simulation results
confirms the predictions of our method.

From an optimization point of view, 
besides the algorithm for discovering it, the most interesting quantity
is the length or energy of the optimum path {\em i.e.} $E_{GS}$; 
thus we now  focus on the zero temperature
limit. First, let us recall that our treatment computes an 
extensive ground state energy. On the other hand, we expect
that for the Euclidean TSP in $d$ dimensions, 
the length of the shortest path should scale as $N^{1-1/d}$
\cite{tspe}.
In this Letter, 
rather than attempt to pursue finite size
corrections to our formalism \cite{inprep},
we study models which do have an extensive ground state energy. 
This is the case in the maximal TSP problem and for problems where 
the potential $V(x)$ between neighboring monomers is repulsive.
We reformulate Eq. (\ref{eqsf})
to work directly at zero temperature by writing 
$s(x) = \exp\left(-\beta w(x)\right)$. Then in the limit $\beta \to \infty$
we find
\begin{equation}
w(x) = {\rm min}_{y\in (0,1)}\left\{ V(x-y) - w(y) \right\}.
\label{eqw}
\end{equation}
The  ground state energy is then given according to 
Eq. (\ref{eqsf}) as 
\begin{equation}
\epsilon_{GS} =2 \int dx \ w(x).
\label{eqgsw}
\end{equation}
The Eq. (\ref{eqw}) seems in general quite difficult to solve, however we can
make progress in some specific cases in one dimension. 
First there is the obvious case of 
what happens when the minimum value of $V(x)$ occurs at $x=0$, as is the
case for the positive temperature TSP. Clearly $\epsilon_{GS} = V(0)$ here
as the optimal solution connects all the sites in ascending (or descending)
order and the distance between each site $\sim 1/N \to 0$. In this case
we see that Eq. (\ref{eqw}) clearly has a solution $w(x) = V(0)/2$ and thus
Eq. (\ref{eqgsw}) implies indeed that $\epsilon_{GS} = V(0)$.

When $V$ is a monotonically decreasing function (so the monomers 
prefer to be as far from each other as possible) a
possible strategy for obtaining the ground state energy is a 
greedy  algorithm where  one starts at the leftmost point 
and goes to the rightmost point, then to the second leftmost point 
and so on. This leads to an energy of   
\begin{equation}
\epsilon_{GA} = \int dx\  V(x).
\label{egreedy}
\end{equation}
Obviously the solution $w(x)$ to Eq. (\ref{eqw}) should be 
symmetric about $x = {1\over 2}$ 
thus $w(x) = u(|x-{1\over 2}|)$ and $u$ obeys
\begin{equation}
u(x) = {\rm min}_{y\in(-{1\over2},{1\over2})}
\left\{ V(x-y) - u(y) \right\}.
\label{equ}
\end{equation}
Inspired by the greedy algorithm we postulate that
the minimization on the right-hand side of  Eq. (\ref{equ}) is achieved 
by $y=-x$. This leads to a solution of Eq. (\ref{equ})  $u(x) = 
{1\over2}V(2x)$, which is valid when $V''(2x) < 0$, {\em i.e} 
when $V$ is concave, for all  $x\in [0,1]$. 
The energy of this solution is given by 
Eq. (\ref{egreedy}) and is thus attained by the greedy algorithm. 
We cannot guarantee that this solution is unique but it is in 
agreement with the simulations for the potentials $V(x) = -|x|$ and $-x^2$ 
\cite{inprep}. 

Another ansatz for solving Eq. (\ref{equ}) is $u(x) = a|x| + b$.
The  Eq. (\ref{equ}) is solved by this form when $V''(1/2) > 0$
(and thus does not work for  concave potentials) and one finds 
 $a= V'({1\over 2})$ and $b = {1\over 2}\left( V({1\over 2}) - 
{1\over2} V'({1\over 2})\right)$. 
The  ground state energy 
predicted by this solution is $\epsilon_{GS} = V({1\over2})$
A potential where the above solution is possible is $V(x)
=-\ln(|x|)$. Numerical solution of Eq. (\ref{eqsff}) at low 
temperatures converges to the solution found above. The 
predicted value of the ground state energy is $\epsilon_{GS} = \ln(2)$, this 
value is compatible with the simulations for this potential
\cite{inprep}. This energy is achieved by making jumps 
from site to site where the jump size is a random
variable $\Delta$ very close to $1/2$ {\em i.e.} if the current position is
$x < 1/2$ one jumps to $ x + \Delta$, otherwise one jumps to $x- \Delta$.
This halfjump-algorithm  will generate an energy per site of $
\epsilon_{HA}= V(1/2)$ and it will also generate the required uniform 
quenched distribution of points on $[0,1]$.
To summarize, the greedy and half-jump algorithms will achieve energies 
per site:
\begin{equation}
\epsilon_{GA} = \int dx\ V(x) ;\ \ \epsilon_{HA} = V({1\over 2}).
\end{equation}
Clearly when $V''(|x|) > 0$
everywhere in $[0,1]$, Jensen's inequality tells us that
$\langle V(X)\rangle \ge V(\langle X\rangle)$ for $X$ distributed on 
$[0,1]$; when this distribution is uniform this implies that
$\epsilon_{GA} > \epsilon_{HA}$ and hence the half-jump algorithm is
the most efficient. In the case where the potential
is concave the greedy algorithm is the most efficient. We note that
the case of the maximal TSP is an intermediate case where 
$V''(x) = 0$ and in this case $ \epsilon_{GA}= \epsilon_{HA}$ and the 
forms of $u(x)$ in these two cases coincide.
If $V(x)$ (for $x >0$) is such that it has a single minimum at
$x=x^*$ and where $x^* < 1/2$, then Eq. (\ref{eqw}) has a solution 
$w(x) = V(x^*)/2$ for all $x$ and the ground state energy 
is $\epsilon_{GS} = V(x^*)$. An algorithm which achieves this
is the  $x^*$-jump  algorithm, note that this works because
when $x^* < 1/2$ the algorithm be applied from any starting
position. When $x^* > 1/2$, $w(x) = V(x^*)/2$ is clearly not a
solution for all $x$ as it fails in the neighborhood of $x=1/2$. 
We note that when $x^* <1/2$ the jumping algorithm   
is in fact a greedy algorithm as it always chooses the near optimal 
jump size and this jump size is independent of its current position. 
When the optimal jump size $x^* > 1/2$ the strategy clearly does not work 
for the reasons discussed above. In this case the  jump sizes 
are no longer concentrated around some typical value and the next jump 
size depends on the current position.  

We have discussed the statistical mechanics of models whose phase space
is the set of permutations of $N$ objects characterized by  
quenched positions/sites. The Hamiltonians are functions of the 
neighboring elements in the sequence, and thus a given sequence can
be interpreted as the energy of a polymer or random walk which visits each
site once. We showed how the quenched calculation 
could be carried out and confirmed its predictions with Monte Carlo 
simulations. Physically the method corresponds to imposing a 
fictitious site dependent chemical potential on the distribution of the set 
of dynamical variables $x_i$ in the presence of the original interaction 
Hamiltonian. This chemical potential is then chosen to ensure that the 
annealed density of these dynamical $x_i$ is the same as the 
desired distribution of the quenched random variables $x_i^{(q)}$. 
The method introduced  works in the thermodynamic limit (corresponding
to high density where the size of the domain is held constant)
for any quenched distribution $p_q(x)$, in any dimension and 
for any interaction potential $V(x)$. For the problem of a directed
polymer in dimensions greater than two a finite temperature phase transition
is known to occur \cite{dego}, it would be interesting to see if this
phase transition shows up in the Euclidean TSP in $d>2$. 
Finally, the idea of treating quenched variables as
effectively annealed variables and then adjusting their Boltzmann  
weight in order to recover, self consistently, the original quenched 
distribution may prove useful  either as an exact or approximate method,
as is the case of Morita's approach \cite{morita,kuhn}, 
in other problems involving quenched disorder.

\pagestyle{plain}

\end{document}